\documentclass[aps,prl,twocolumn,superscriptaddress,showpacs]{revtex4}
\usepackage{graphicx}

\begin{document}

\title{Aharonov-Bohm Effect in Concentric Quantum Double Rings }
\author{Guang-Yin Chen}
\affiliation{Institute of Physics, National Chiao Tung University, Hsinchu 300, Taiwan}
\author{Yueh-Nan Chen}
\email{ynchen.ep87g@nctu.edu.tw}
\affiliation{Department of Physics, National Cheng Kung University, Tainan 701, Taiwan}
\affiliation{National Center for Theoretical Sciences, Tainan 701, Taiwan}
\author{Der-San Chuu}
\affiliation{Department of Electrohysics, National Chiao Tung University, Hsinchu 300,
Taiwan}
\date{\today}

\begin{abstract}
We propose a theoretical model to study the single-electron spectra of the
concentric quantum double ring fabricated lately by self-assembled
technique. Exact diagonalization method is employed to examine the
Aharonov-Bohm effect in the concentric double ring. It is found the
appearance of the AB oscillation in total energy depends on the strength of
the screened potential. Variations of the energy spectra with the presence
of coulomb impurities located at inner or outer ring are also investigated.
\end{abstract}

\pacs{73.22.-f}
\maketitle





Due to the presence of ring-like structures in the past two decades,
observing Aharonov-Bohm (AB) effect$^{1}$ is no longer impossible. The first
observation of AB effect in normal metal rings was reported by Webb \emph{et
al.$^{2}$}. On the theoretical side, Cheung \emph{et al.$^{3}$ } calculated
the persistent currents and energy levels of the electron in a
one-dimensional ring. Gap-like structure was found with the presence of
impurities.

In addition to the metallic quantum rings, A. V. Chaplik$^{4}$ in 1995
considered a semiconducting antidot under a strong magnetic field. It was
found the tunnelings of the electron and hole around the antidot may result
in a shift for each excitonic level. R\"{o}mer and Raikh$^{5}$ then found
similar results by using a quite different analytical approach. With the
advances of self-assembled techniques, nanoscopic semiconductor rings were
fabricated recently with a characteristic inner/outer radius of 10/30-70 nm
and 2-3 nm in height $^{6,7}$. After the successful experimental
realizations, additional effects, such as finite width$^{8,9}$, presence of
barriers$^{10}$, and penetration of the magnetic field into ring region$%
^{11} $, on the excitonic levels were also taken into account theoretically.

Very recently, Mano \emph{et al.$^{12}$ } successfully demonstrated the
formation of concentric quantum double rings with high uniformity and
excellent rotational symmetry. The diameters of the inner and outer rings
are 45 ($\pm 3$)nm and 100 ($\pm 5$)nm, respectively. Electronic structures
of the concentric quantum double rings are then calculated theoretically.$%
^{13}$. Motivated by these reports, we thus propose in this work a
theoretical model to study the AB effect in concentric quantum double rings.
The behavior of the AB oscillations in the double ring will be shown to
depend on the positions and strengths of the coulomb impurities.

\begin{figure}[tbp]
\scalebox{0.45}{\includegraphics*[8mm,89mm][205mm,285mm]{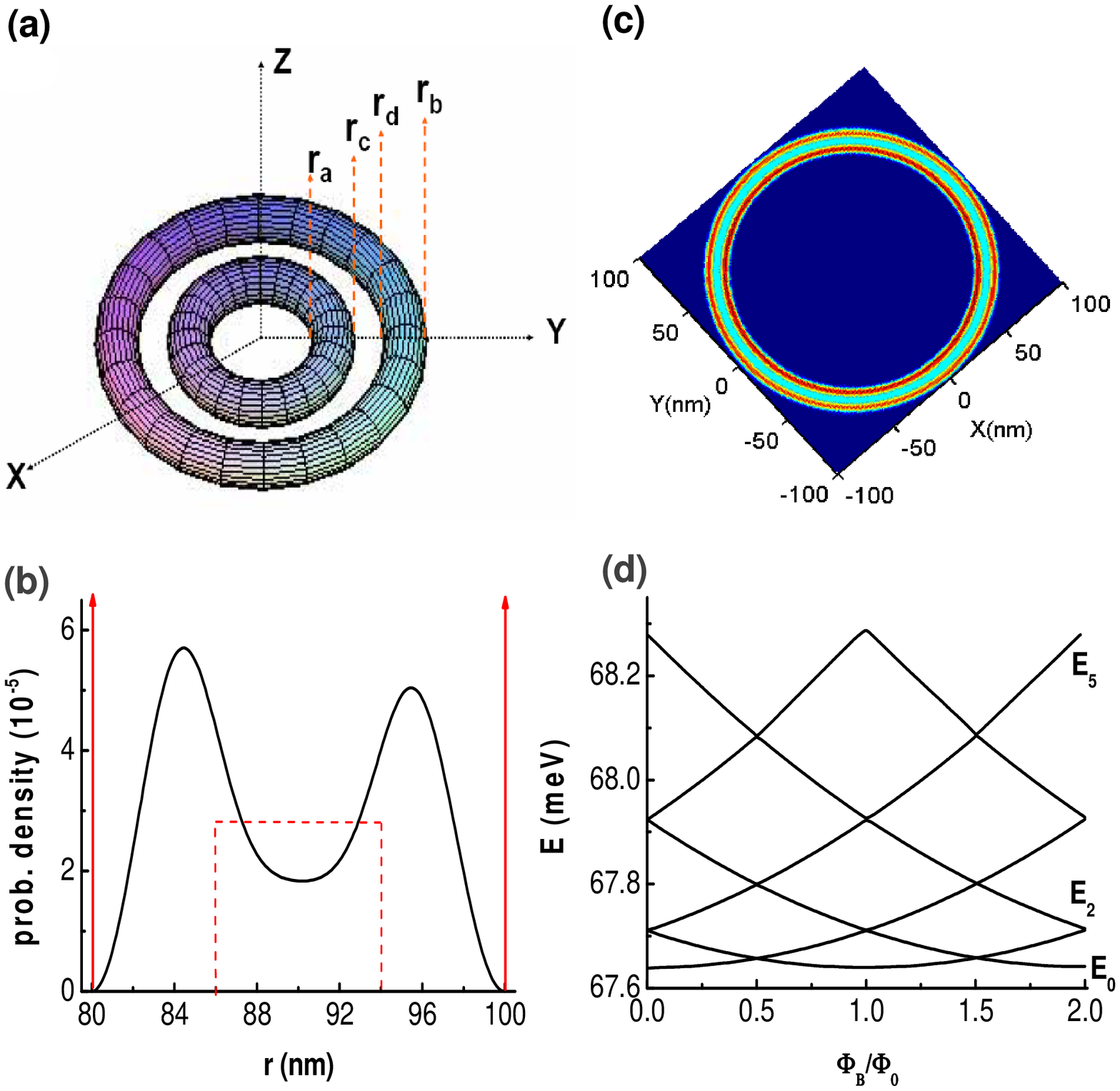}}
\caption{(a)Illustration of the concentric quantum double ring. (b) Model
potential of the concentric quantum double ring and the corresponding ground
state wavefunction for an electron in the radial direction. (c) The
probability density of the ground state electron. (d) Energy levels as a
function of the magnetic flux $\Phi _{B}$ threading through the interior of
the double rings.}
\end{figure}

\emph{The model.} ---Consider now an electron in a two-dimensional
concentric double ring with the inner ring shell enclosing a uniform static
magnetic flux ($\Phi _{B}=B\pi r_{a}^{2}$) oriented along the z axis. The
concentric double ring structure is approximated by the hard-wall
confinement potential with a tunable constant barrier in the middle as shown
in Fig. 1 (b). The model Hamiltonian can be written as
\begin{equation}
H=\frac{(\vec{P}+\frac{e\vec{A}}{c})^{2}}{2m_{e}^{\ast }}+V(r),
\end{equation}%
where $\vec{P}$ is canonical momentum, $\vec{A}$ is the magnetic vector
potential, and $m_{e}^{\ast }$ is the effective mass of the electron. The
potential $V$ is then defined as
\begin{equation}
V(r)=\cases{ \infty ,& if $r< r_a , r > r_b$ \cr V_0,& if $r_c < r < r_d$
\cr 0 & otherwise \cr}.
\end{equation}%
The Hamiltonian operator inside the ring is
\begin{equation}
\hat{H}=\frac{\hat{P_{r}}^{2}}{2m_{e}^{\ast }}+\frac{(\hat{P_{\phi }}+\frac{%
eA_{\phi }}{c})^{2}}{2m_{e}^{\ast }}+V_{0}.
\end{equation}%
The vector potential needed to produce the magnetic field $\vec{B}$ $(=B\hat{%
z})$ in the interior of the double ring is $\vec{A}=\frac{Br_{a}^{2}}{2r}%
\hat{\phi}$. The position representation of Eq. (3) can be written as
\begin{equation}
H=-\frac{\hbar }{2m_{e}^{\ast }}(\frac{d^{2}}{dr^{2}}+\frac{1}{r}\frac{d}{dr}%
)-\frac{\hbar ^{2}}{2m_{e}^{\ast }r^{2}}(\frac{\partial }{\partial \phi }+%
\frac{e\Phi _{B}}{2\pi \hbar c})^{2}+V_{0}.
\end{equation}%
As is well known, the wave function $\psi _{n,m}(r,\phi )$ ($n,m$ denote the
principal and the magnetic quantum number, respectively) of the electron in
the single ring is given by
\begin{equation}
\psi _{n,m}(r,\phi )=[C_{1}J_{|m|}(kr)+C_{2}Y_{|m|}(kr)]\cdot (\frac{1}{%
\sqrt{2\pi }}e^{im\phi }),
\end{equation}%
where $C_{1}$ and $C_{2}$ are the normalization constants. To solve the
double ring problem, $\psi _{n,m}$ is chosen as the complete set to span the
Hamiltonian operator. The energy eigenvalues $E_{j}$ and the corresponding
eigenfunctions $|\Phi _{i}\rangle $ can then be obtained via numerical
diagonalization of $\hat{H}$ in the matrix representation. In our
calculations, the effective Rydberg unit, $Ry^{\ast }=m_{e}^{\ast
}e^{4}/2\hbar ^{2}(4\pi \epsilon _{0}\epsilon _{r})^{2}=5.8meV$, is adopted.
The unit in length is effective Bohr radius, $a_{B}^{\ast }=4\pi \varepsilon
_{0}\varepsilon _{r}\hbar ^{2}/m_{e}^{\ast }e^{2}=10nm$. The effective mass
of the electron is $0.067m_{e}$, and the universal flux quanta $\Phi _{0}$
for a radius of $a_{B}^{\ast }$ corresponds to a magnetic field of $13.18T$
in $GaAs/Al_{0.3}Ga_{0.7}$. The boundaries of the double rings are chosen as

\begin{equation}
\left\{
\begin{array}{c}
r_{a}=80nm(8a_{B}^{\ast }) \\
r_{b}=100nm(10a_{B}^{\ast }) \\
r_{c}=86nm(8.6a_{B}^{\ast }) \\
r_{d}=94nm(9.4a_{B}^{\ast })%
\end{array}%
\right.
\end{equation}%
with the middle barrier $V_{0}=100meV$ , which makes the tunneling between
two rings become possible.

Fig. 1. (c) and (d) show respectively the probability density of the ground
state and the energy spectra of the electron in the presence of magnetic
flux $\Phi _{B}$. Just like the behavior in single ring, the AB effect is
clearly seen whenever $\frac{\Phi _{B}}{\Phi _{0}}$ ($\Phi _{0}\equiv \frac{%
hc}{e}$ is the universal magnetic quantum) is a positive integer.
\begin{figure}[tbp]
\scalebox{0.6}{\includegraphics*[30mm,125mm][180mm,270mm]{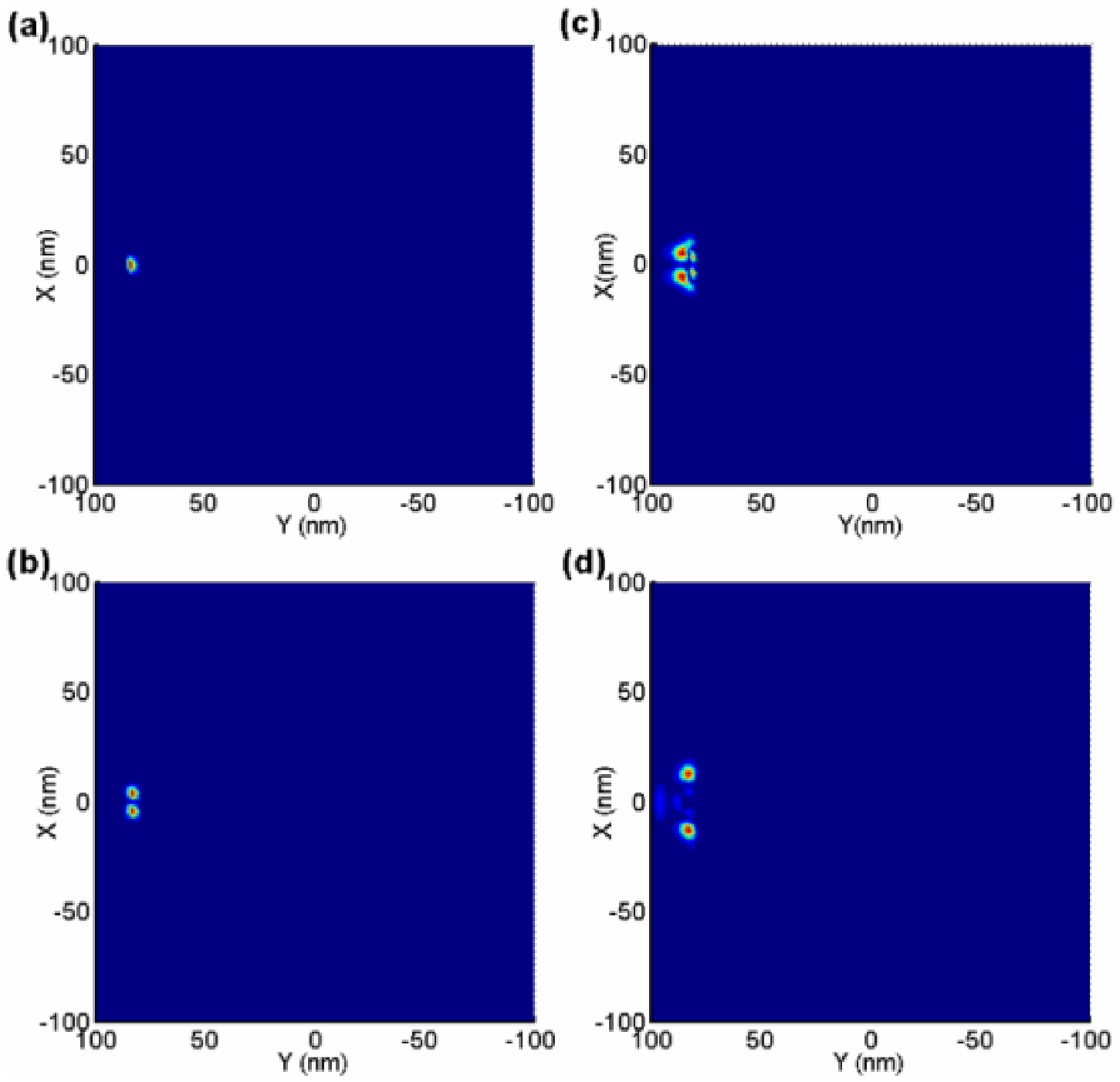}}
\caption{Density plot of the electron wavefunction in the (a) ground state ($%
E_{0}$); (b) second excited state ($E_{2}$); (c) fifth excited state ($E_{5}$
); (d) ninth excited state ($E_{9}$).}
\end{figure}

\emph{Effects of fixed impurities.} ---Let us now consider a positively
charged impurity located at the inner ring $(r_{i}=83nm,\phi _{i}=0)$ of a
concentric quantum double ring. The potential between the electron and
impurity can be described by the screened coulomb (Debye) potential$^{14}$
\begin{equation}
U=\frac{-e^{2}\cdot e^{-\alpha |\vec{r}_{e}-\vec{r}_{i}|}}{\epsilon _{r}|%
\vec{r}_{e}-\vec{r}_{i}|},
\end{equation}%
where the $\epsilon _{r}$ is the static dielectric constant (for GaAs, $%
\epsilon _{r}=12.4$) and $\alpha $ is the screening parameter. Let us first
consider the case of bared coulomb potential, i.e. $\alpha =0$. Probability
densities of the electron for various quantum states are shown in Fig. 2. As
can be seen, even for the ninth excited state ($E_{9}$), the wavefunction
that penetrates into the outer ring is still few.

Unlike the AB behavior in Fig. 1, the total energy ($E_{t}$) of the electron
in the low quantum state increases monotonically with the increasing of the
magnetic flux as shown in Fig. 3 (a) and (b). The reason is that the bared
coulomb potential is so strong, such that it's almost impossible for the
electron to circle around the ring. However, for the case without the
impurity, the kinetic energies ($E_{k}$) in Fig. 1 do show the AB
oscillation. From the definition of binding energy, $-E_{b}=E_{t}-E_{k}$,
the AB oscillation can be revealed as one plots the variations of binding
energy in Fig. 3 (c) or (d). One also notes that the binding energy of the
second excited state ($E_{2}$) increases as the flux is increased from zero
to half of $\Phi _{0}$. On the contrary, the binding energy of the fifth
excited state ($E_{5}$) decreases first in the beginning. This tendency can
also be understood from the\ combination of the two factors: the oscillatory
kinetic energies in Fig. 1 (d) and the monotonically increasing behavior in
Fig. 3 (a) and (b).
\begin{figure}[tbp]
\scalebox{0.45}{\includegraphics*[8mm,60mm][199mm,274mm]{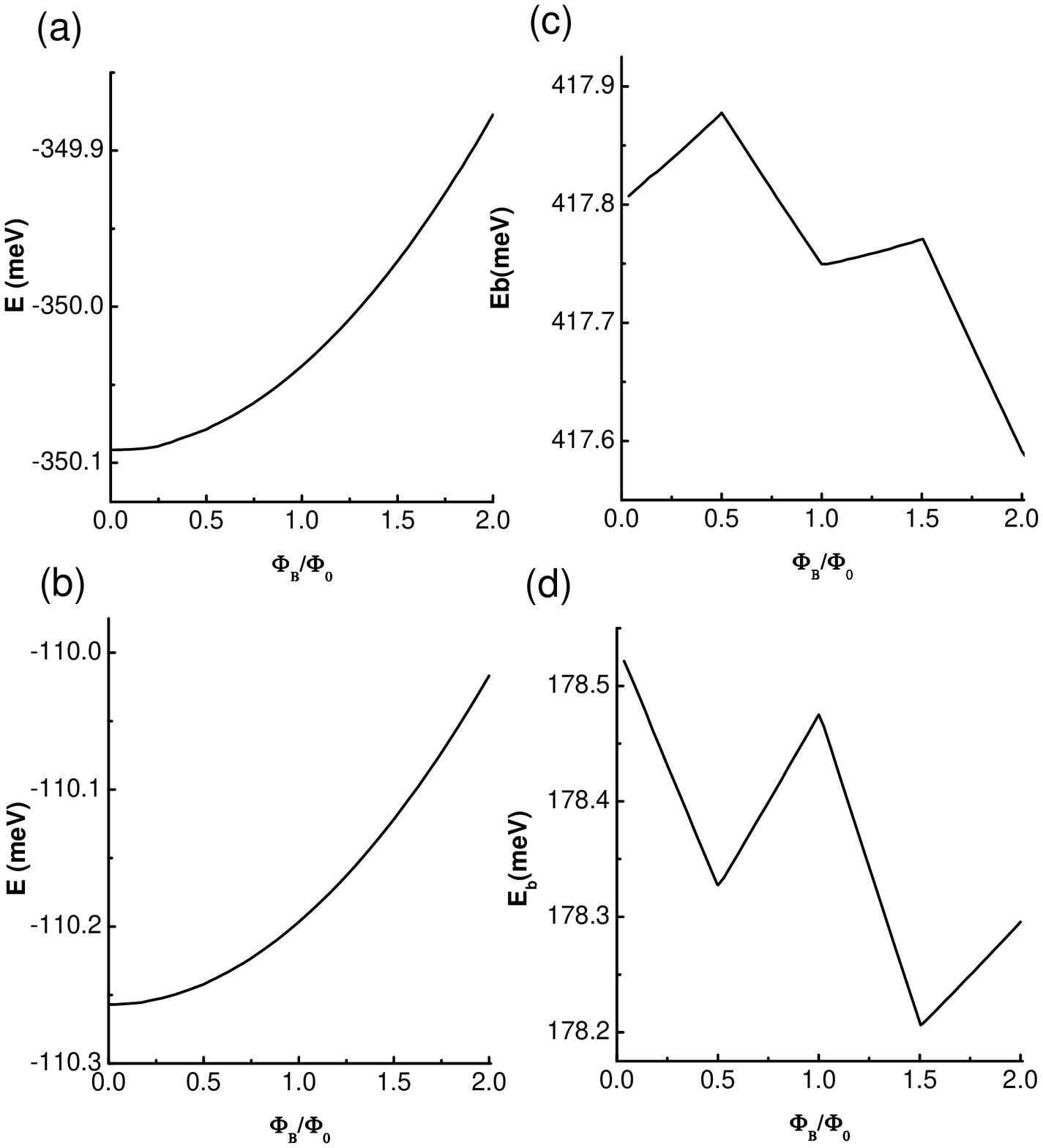}}
\caption{(a) and (b) represent the total energies ($E_{t}$) of the second
and fifth excited states, respectively. (c) and (d) show the corresponding
binding energies ($E_{b}$), which exhibit AB oscillations, as functions of $%
\Phi _{B}$.}
\end{figure}

To see the effect of the screening parameter, it is convenient to define the
parameter $\delta =\alpha a_{B}^{\ast }$, which contains no unit. Fig. 4
represents the probability densities of the ground state wavefunctions for
different values of $\delta $. As can be seen, the probability density of
the electron gathered around the position of the impurity when $\delta =0.05$%
. As $\delta $ increases, the probability density of the electron begins to
spread in the concentric quantum double ring. Since we are interested in the
problem of impurity mixed with AB effect, we choose $\delta =0.4$, for which
the\ wavefunction spreads but not too much, in the rest of the discussions.

\begin{figure}[tbp]
\scalebox{0.6}{\includegraphics*[30mm,128mm][179mm,272mm]{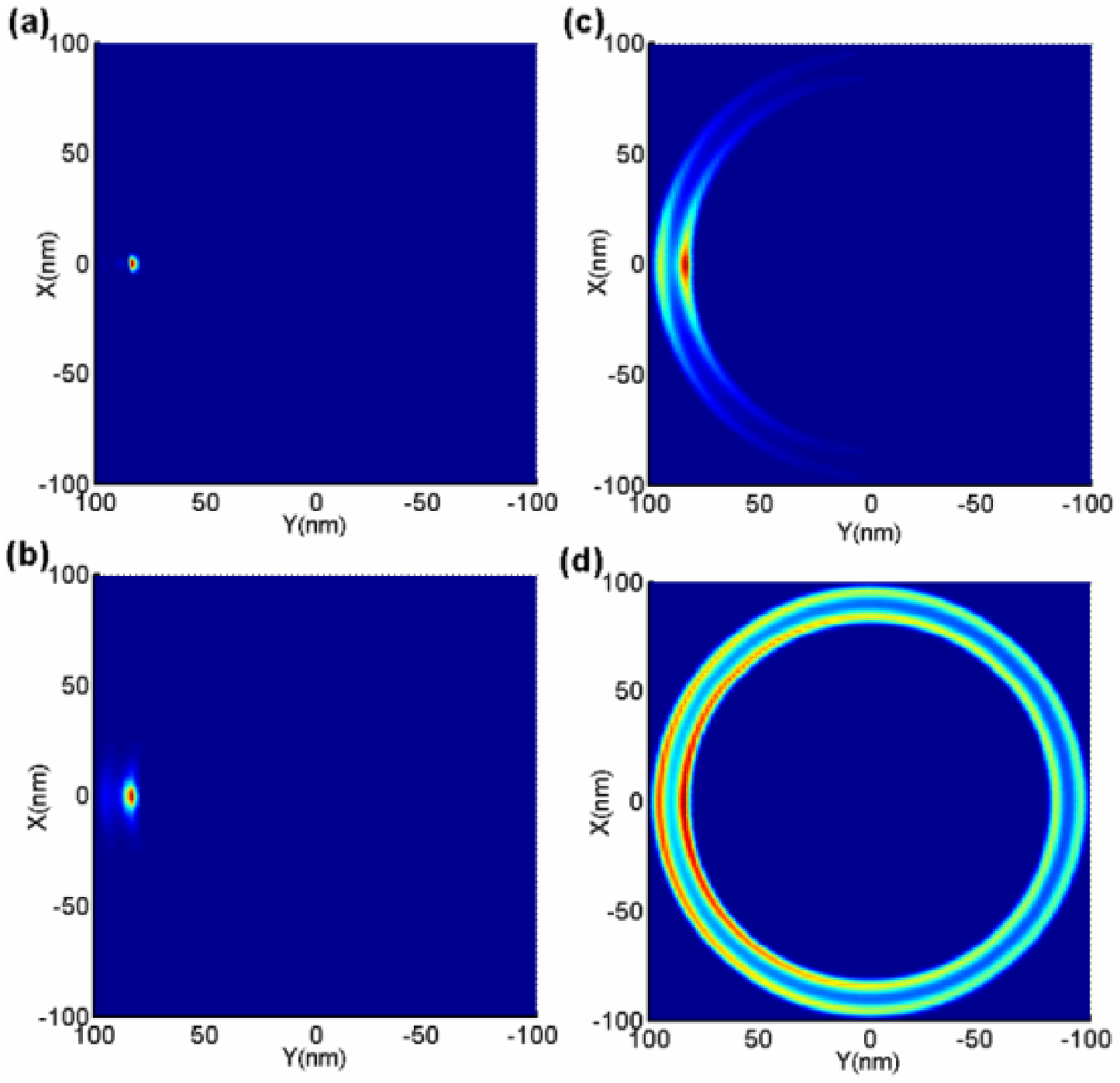}}
\caption{Probability density of the ground state electron with a
fixed impurity located at $(r_{i}=83nm,\protect\phi _{i}=0)$ in the
concentric quantum double ring with (a) $\protect\delta =0.05$, (b)
$\protect\delta =0.2 $, (c) $\protect\delta =0.4$, and (d)
$\protect\delta =0.6$.}
\end{figure}

The energy spectra for the uniform magnetic flux threading the interior of
the concentric double ring are shown in Fig. 5 (a). Since the coulomb
potential is much weaker$^{14}$, the AB effect can now be revealed in the
spectra of total energy. Also, due to the fixed impurity, the cylindrical
symmetry of the concentric double ring is broken and the two-folded energy
degeneracies for each $m$ ($m\neq 0$) are then removed. As a result, the AB
effect is reduced in the presence of the fixed impurity. If one compares
Fig. 5 (a) with Fig. 1 (d), the gaps are found to be opened at the points of
intersecting curves in Fig. 1. (d), just like the band structures in solid
state physics. This is because the magnetic flux, $2\pi \Phi _{B}/\Phi _{0}$%
, plays the same role as $kL$ in one-dimensional Bloch problem$^{15}$, where
$k$ is the wave vector and $L$ is the length of the one dimensional lattice.
The energy levels of the ring-formed microbands as a function of $\Phi _{B}$
are analogous to the Bloch electron bands in the extended $k$ -zone picture.

If one further puts an additional impurity which is opposite to the first
impurity in the inner ring, the degeneracies are removed further as shown in
Fig. 5 (b), i.e. the separations of originally degenerate states (for
example, $E_{7}$ and $E_{8}$) become bigger. However, since two impurities
are symmetric, it also forces the degenerate states to intersect with
another states (for example. $E_{6}$ and $E_{7}$). That's why the ''energy
gaps'' look like disappear. The most important characteristic for concentric
quantum rings is the inner and outer ring structure. One can move the second
impurity to the middle of outer ring ($r=97nm,\phi =\pi $). As shown in Fig.
5 (c), the energy gaps are opened again. The reason is that the wave
functions are not symmetric with respect to the middle barrier $V_{0}$ [see
Fig. 1. (c)]. Therefore, the attractive ability of the two impurities
located at different rings are not the same. In addition, one can also
suppress the azimuthal symmetry, i.e. locating the two impurities at the
same ring (inner ring) but at different azimuthal angles $(\phi =0,\frac{\pi
}{2})$. As can be seen from Fig. 5 (d), the energy separations are larger
than those for single impurity case, and the curves do not intersect with
each other.

\begin{figure}[tbp]
\scalebox{0.45}{\includegraphics*[10mm,15mm][193mm,267mm]{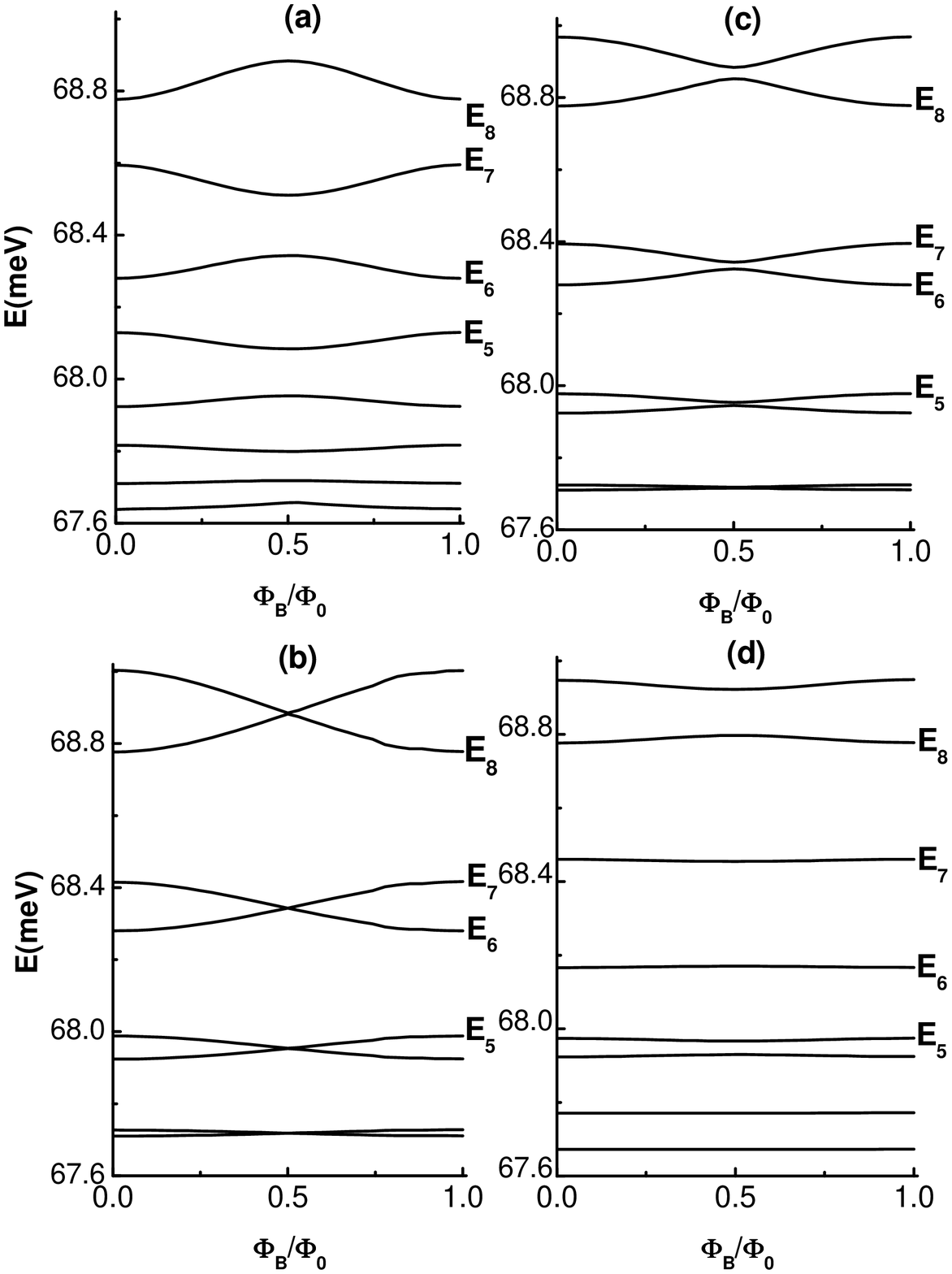}}
\caption{Energy spectra as a function of the magnetic flux $\Phi _{B}$ with
(a) a fixed impurity located at $(r_{i}=83nm,\protect\phi _{i}=0)$; (b) two
symmetric fixed impurities located at $(r_{i1}=83nm,\protect\phi _{i1}=0)$
and $(r_{i2}=83nm,\protect\phi _{i2}=\protect\pi )$; (c) two fixed
impurities located at $(r_{i1}=83nm,\protect\phi _{i1}=0)$ and $(r_{i2}=97nm,%
\protect\phi _{i2}=\protect\pi )$; (d) two asymmetric fixed impurities
located at $(r_{i1}=83nm,\protect\phi _{i1}=0)$ and $(r_{i2}=83nm,\protect%
\phi _{i2}=\frac{\protect\pi }{2})$.}
\end{figure}

In summary, we have studied the AB effect in a concentric double ring.
Effect of impurities with screened coulomb potential on the energy spectra
is studied. The appearance of the AB oscillation in total energy is found to
depend on the strength of the screened potential. In the case of screened
potential, the openings and closings of the energy gaps depend not only on
the angular distributions of the impurities, but also on whether the
impurities are located at the same ring, which is the unique feature for the
double ring systems.

\subsection{\textbf{ACKNOWLEDGMENTS}}

We thank Prof. Johnson Lee at Chung Yuan Christian University and Dr. W. H.
Kuan at Center for Theoretical Sciences for helpful discussions. This work
is supported partially by the National Science Council, Taiwan under the
grant number NSC 95-2119-M-009 -030.

\end{document}